# Computational insights into Cobalt-based novel half-Heusler alloy for sustainable energy applications

Sumit Kumar[a,b], Diwaker[c], Ashwani Kumar[d], Vivek[a], Arvind Sharma[e], Karan S. Vinayak[b,**] and Shyam Lal Gupta[f,*]

[a]Department of Physics, Government College Una, H P , 174303, INDIA
[b]Department of Physics, DAV College Sec-10, Chandigarh, 160010, INDIA
[c]Department of Physics, SCVB Government College, Palampur, Kangra, 176061, H P , INDIA
[d]School of Basic Sciences, Abhilashi University Mandi, Mandi, 175045, H P , INDIA
[e]Department of Physics, SVSDPG College Bhatoli, Una, 174301, H P , INDIA
[f]Department of Physics, HarishChandra Research Institute, Prayagraj, Allahabad, 211019, U P , INDIA

ABSTRACT

The quest for efficient and sustainable green energy solutions has led to a growing interest in half-Heusler alloys, particularly for thermoelectric and spintronic applications. This study investigates the multifaceted nature of cobalt-based half-Heusler alloy, CoVAs, employing density functional theory (DFT) with advanced computational techniques, such as the full-potential linearized augmented plane wave (FLAPW) method. The elastic, electronic, magnetic, thermodynamic, and optical properties of CoVAs are meticulously analyzed. Structural and mechanical evaluations reveal mechanical stability and brittleness under varying pressures. Electronic and magnetic properties are examined through band structure and density of states (DOS) analyses, revealing a half-metallic nature with a minority-spin band gap of 1.174 eV. The total magnetic moment of 1.0000 B aligns with the Slater-Pauling rule, further confirming ferromagnetism and half-metallicity. Thermodynamic investigations, based on the quasi-harmonic Debye approximation, provide insights into temperature- and pressure-dependent behavior, including thermal expansion, heat capacity, and Debye temperature, establishing CoVAs as a viable candidate for high-temperature applications. Additionally, the optical properties underscore its potential in optoelectronic applications due to high absorption in the UV region, showing a distinct absorption edge corresponding to the electronic band gap. Phonon dispersion relations refelct the stability of the alloy, while the figure of merit(ZT= 0.74) confirms the alloy's suitability for thermodynamics applications. The findings highlight the potential of CoVAs as a promising candidate for spintronic photovoltaic and optoelectronic applications, providing insights into its fundamental properties that could facilitate experimental synthesis and industrial implementation for green energy and advanced technological applications

## 1. Introduction

The transition to sustainable energy technologies has spurred intense research into novel materials capable of delivering high efficiency and performance. Half-Heusler (HH) alloys have emerged as versatile candidates for green energy applications, owing to their unique structural, electronic, and thermoelectric properties [1, 2, 3, 4, 5, 6, 7, 8, 9, 10, 11]. Among these, cobalt-based half-Heusler alloys have garnered attention due to their tunable magnetic and electronic behaviors, making them suitable for thermoelectric devices, spintronic systems, and optoelectronic applications [12, 13]. Heusler alloys are broadly classified into two categories based on their structures: full-Heusler alloys and half-Heusler alloys, denoted as $X_2YZ$ and $XYZ$, respectively. In these compounds, X and Y belong to transition metals, while Z can be either a main group element or a transition element( all d-metal [14]). Moradi et al. [15] investigated the half-metallic properties of the half-Heusler NaZrZ (Z = P, As, Sb) alloys and anlyzed the impact of pressure on these characteristics. Davatolhagh and associates explores the behaviour of half-metallic materials by replacing the $d^0$ alkaline metal atoms in traditional HH with low valence transition metals [16]. Their study demonstrated that the half-metallic nature of these materials could be attributed to the $\beta$ phase of HH alloys, which consist of a combination of $d^0$ alkali metals and 3d transition metals. T. Zerrouki et al. [17] employed the full potential linear muffin-tin orbital method with the generalized gradient approximation (GGA) and (GGA+U) approach to study the physical properties of NbCoSn and NbFeSb Hh alloys. Their findings confirmed the mechanical stability of these materials and established their semiconductor nature. Additionally , M. H. Elahmar et.al [18] used DFT simulations to examine the thermoelectric properties of novel HH NbFe(Mn)Sn compounds and their layered superlattices along the [100] direction. A key factor in predicting the fundamental properties of Heusler alloys including superconductivity, electrical conductivity, semiconducting behaviour, and magnetism is the number of valence electrons [19].For instance, full Heusler alloys with 27 valence electrons and no magnetic characteristics exhibit

*Corresponding author
**Corresponding author

drksvinayak.dav@gmail.com (S. Kumar); diwakerphysics@gmail.com ( Diwaker); vjairath27@gmail.com ( Vivek); arvindkamalsharma@gmail.com (A. Sharma); drksvinayak.dav@gmail.com (K.S. Vinayak); shyamlalgupta@hri.res.in (S.L. Gupta)

ORCID(s): 0000-0001-7276-8287 (S. Kumar); 0000-0002-4155-7417 ( Diwaker); 0000-0003-0472-3436 (S.L. Gupta)

superconducting behaviour, while full Heusler alloys with 24 valence electrons display semiconducting characteristics. Furthermore, half-Heusler alloys with 8 or 18 valence electrons exhibit remarkable thermoelectric and optical properties and are classified as ferromagnetic semiconductors [20, 21, 22, 23, 24, 25, 26, 27, 28, 29]. The objectives of this study are threefold: (1) To assess the mechanical stability of coVAs under varying pressure and determine its brittleness or ductility. (2) To explore its electronic and magnetic properties to understand its half-metallic behaviour. (3) To investigate its thermodynamics, thermoelectric, and optical responses to evaluate its potential for technological applications. In this work, we focus on the 19- valence electrons CoVAs HH alloy, employing the density functional theory (DFT) framework to gain a comprehensive understanding of its properties. The rest of the paper is structured as follows: section 2 details the computational methodology. Section 3 presents our first-principles investigation results, covering structural and mechanical properties first, followed by electronic and magnetic properties, thermodynamic properties, and optical properties. additionally , we evaluate the stable structural phase through lattice dynamics studies and assess the figure of merit. To the best of our knowledge, the half-metallic properties of CovAs HH alloy have not been previously reported in the literature.

## 2. Computational details

The calculations presented in this paper utilize the full-potential linearized augmented plane wave (FLAPW) method, within the framework of density functional theory (DFT)[30], [31], [32] as implemented in WIEN2k simulation package[33], to investigate the multifaceted nature of the CoVAs heusler alloy. FLAPW method proves to be a dependable option, given the sensitivity of certain Heusler alloys to the chosen method[12]. The generalized gradient approximation(GGA) scheme serves as the exchange-correlation potential for lattice constant calculations. However, due to GGA's inherent underestimation of band profiles resulting from strongly correlated mechanism, the modified Becke-Johnson (mBJ) potential is employed to enhance accuracy[34, 35, 36, 37]. The structural unit cell is divided into two regions to initiate the simulations: the interstitial space and the muffin-tin sphere. The Muffin-Tin approximation is a widely used and highly compatible method for determining the energy states of an electrons within a crystal lattice. Furthermore, elastic constants, defined as second derivatives of total energy concerning lattice deformation, are determined using the IRelast package within WIEN2k, facilitating comprehensive calculations of elastic constants and related parameters. Thermoelectric properties of the alloy are explored using the BoltzTraP2 code[38]. The exchange -correlation potential which is considered for band strucutre and density of states (DOS) is given as

$$V_{X,\sigma}^{mBJ}(r) = CV_{X,\sigma}^{mBJ}(r) + (3C-2)\frac{1}{\pi}\sqrt{\frac{5}{12}}\sqrt{\frac{2t_{\sigma}(r)}{\rho_{\sigma}(r)}} \quad (1)$$

where $t_{\sigma(r)} = \frac{1}{2}\sum_{i}^{N_\sigma} \nabla_{\psi_{i,\sigma}}$ represents the kinetic energy and $\rho_{\sigma(r)} = \sum_{i}^{N_\sigma} \psi_{i}^{N_\sigma}.\psi_{i,\sigma}$ as the charge desnity. The conventional unit cell is divided into non overlapping spheres with spherical wave function and into an interstitial region possessing plane-wave character. The limit for energy convergence is set to $10^{-6}$ Ry and for the plane wave expansion RMT×$K_{MAX}$ was chosen equal to 9. Here $K_{max}$ is the cutoff for the wave functon basis and RMT is the smallest radii of the muffin-tin spheres. For integration of the Brilion zones (BZ), we used a mesh of 1000 k-points along a grid of 10 × 10 × 10 and cutoff energy which usually seperates the valence and the core states was chosen around -8.0 Rydberg. The minimum selected values of RMT radii are taken as Co =2.20, V =2.09 and As = 2.09. The ground state energy for the ferromagnetic (FM) and non-magnetic (NM) phases is examined in $C1_b$ structures. The elastic parameters of the alloy under investigation are determined using IRElast code [39] within WIEN2k package, while the thermodynamic properties are analyzed using the Gibbs2 code[40]. Furthermore, the percentage of spin polarization (SP) is calculated as given by[41], [42].

$$SP = \frac{\eta^\uparrow - \eta^\downarrow}{\eta^\uparrow + \eta^\downarrow} \times 100 \quad (2)$$

where SP is the percentage of the difference in the states of majority spin($\eta^\uparrow$) and minority spin($\eta^\downarrow$) in ratio.

## 3. Result and Discussion

This section will discuss the properties of the CoVAs half-Heusler alloy in detail across multiple subsections.

### 3.1. Structural and mechanical properties

The physical properties of any material can only be understood by analyzing and visualizing its crystal structure. A full Heusler alloy relaxes in centrosymmetric cubic FCC with structure type $L2_1$ (Fm-3m space group no. 225), whereas the half Heusler (HH) alloys mostly crystallize in non-centrosymmetric cubic FCC with structure type $C1_b$ (F-43m space group no. 216). For half-Heusler alloys, there are three possible structure type: type $\alpha$, with Co at (1/4, 1/4, 1/4), V at (1/2, 1/2, 1/2), and As at (0, 0, 0); type $\beta$, with Co at (1/2, 1/2, 1/2), V at (1/4, 1/4, 1/4), and As at (0, 0, 0); and type $\gamma$, with Co at (0, 0, 0), V at (1/2, 1/2, 1/2), and As at (1/4, 1/4, 1/4). These three types are distinguished by their distinct atomic coordinates and Wyckoff positions. Table-1 lists the exact Wyckoff positions for type $\alpha$, type $\beta$, and type $\gamma$ in the CoVAs alloy.

The computed results exhibited that the ferromagnetic (FM) phase type $\alpha$ of CoVAs crystal structure is most stable in comparison to $\beta$ and $\gamma$ phase types. We observed that the dynamical stability of the proposed system more or less depends upon the atomic positions in the given crystal structure. We have calculated the lowest energy and mean lattice parameters for the cubic CoVAs system with stable type $\alpha$ crystal structure, as shown in Table 2. The Murnaghan equation of the states [44] for volume optimization in the ferromagnetic state is successfully implemented, as expressed in Eqn. 1.

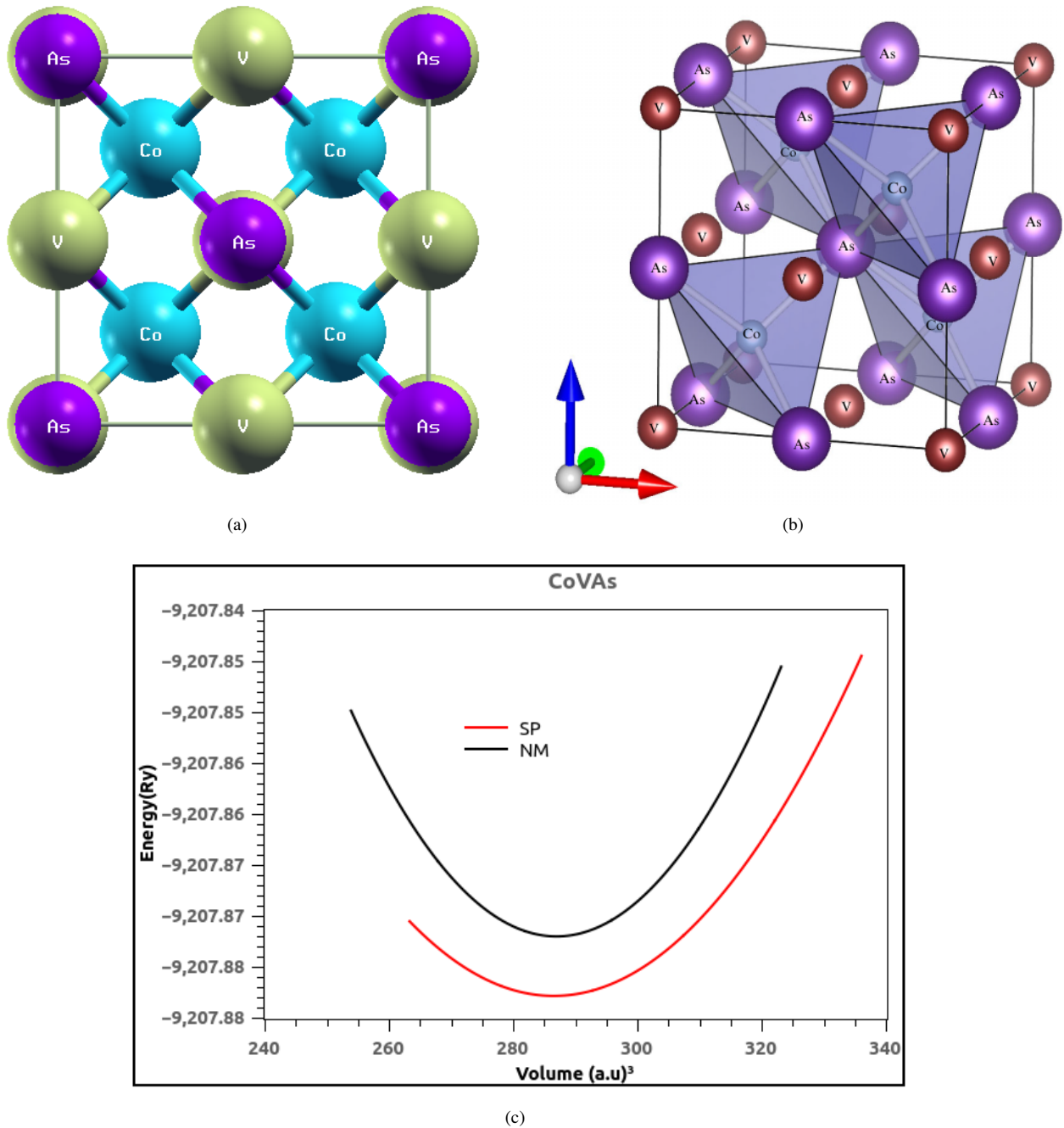

Figure 1: [a] Optimized unit cell [b] CoVAs crystal structure [43] [c]Energy versus Volume curve of CoVAs HH alloy

$$E_T(V) = E_o + \frac{B_o V}{B'_o}\left(\frac{(\frac{V_o}{V})^{B'_o}}{B'_o - 1} + 1\right) - \frac{B_o V o}{B'_o - 1} \quad (3)$$

where $V_0$ represent the unit cell's volume at zero pressure, $B_0$ is the bulk modulus and $E_0$ signifies the minimal total energy. The computed lattice constant ($a^0$) with GGA scheme for CoVAs HH alloy at symmetry alongwith the other parameters $V_0$, $B_0$, and $B'_0$ are reported for the first time in Table 2.

The mechanical and elastic behaviour of CoVAs HH alloy can be evaluated by calculating its unique elastic parameters and analyzing the relationship between stress and strain. The elastic parameters control how the alloys deform when subjected to an external force, allowing the material to revert to its original configuration once the force is removed. Examining the mechanical stability of alloys requires an understanding of the elastic constants under ideal circumstances. In order to determine the use of alloys in mechanical applications, the elastic constants are calculated by using the IRElast code. We calculated the values of the three distinct elastic parameters, $C_{11}$, $C_{12}$, and $C_{44}$, from the calculated data for the phase $\alpha$ cubic symmetry. $C_{12}$ measures the transverse deformation, while $C_{11}$ evaluates the longitudinal expansion. The equilibrium elastic coefficients largely meet Born's stability requirements to characterize the mechanical stability which are $C_{11} > 0$, $C_{44} > 0$, $(C_{11}-C_{12}) > 0$, $(C_{11}+2C_{12}) > 0$. Table-1 depicts the mechanical stability and shows that the CoVAs HH alloy meet the stability requirements. Additionally, the two parameters mentioned as bulk modulus (B) and shear modulus (G), respectively determines the stiffness and compressibility. The parameters G and B can be used to calculate the restoring force which comes into play against the fracture, or resistance to plastic deformation. The following equations can be used to get various parameters. The bulk modulus, B is given as

$$B = \frac{c_{11} + 2c_{12}}{3} \quad (4)$$

also, the shear modulus is given as

$$G = \frac{G_R + G_V}{2} \quad (5)$$

**Table 1**
Wyckoff positions and three potential phase type $\alpha$, $\beta$ and $\gamma$

| Structure | Co | V | As |
|---|---|---|---|
| $\alpha$ | (0.25,0.25,0.25) | (0.5,0.5,0.5) | (0,0,0) |
| $\beta$ | (0.5,0.5,0.5) | (0.25,0.25,0.25) | (0,0,0) |
| $\gamma$ | (0,0,0) | (0.5,0.5,0.5) | (0.25,0.25,0.25) |

**Table 2**
Calculated crystal structure stability parameters.

| Alloy | Type | a0(Å) | B0(GPa) | V (a.u.3) | E(Ry)) |
|---|---|---|---|---|---|
| | | FM | FM | FM | FM |
| CoVAs | $\alpha$ | 5.54 | 169.946 | 287.0010 | -9207.879058 |
| | $\beta$ | 5.68 | 122.09 | 309.4737 | -9207.796676 |
| | $\gamma$ | 5.61 | 150.55 | 298.5040 | -9207.809004 |

The R and V in the subscript of eqn. 3 and eqn 4. denotes the Reuss and Voigt bounds and their values are determined by using the subsequent relations given as

$$G_V = \frac{1}{5}(c_{11} - c_{12} + 3c_{44}) \quad (6)$$

and

$$G_R = 5\left[\frac{(c_{11} - c_{12})c_{44}}{3(c_{11} - c_{12}) + 4c_{44}}\right] \quad (7)$$

The mathematical link between the various elastic constants is consistent with the cubic crystal symmetry. The internal strains associated with bond twisting and elongation are computed using the Kleinmann parameter ($\xi$). It assesses how a basic link would bend in the presence of external forces. The CoVAs HH alloy increased resistance to a wider range of stresses is indicated by the computed value of ($\xi$), confirming their wider usage in industrial applications. The thermodynamic stability, specific heat at low temperature, and phonon stability were confirmed by calculating the Debye temperature ($\theta_D$) using the average sound velocity. Debye temperature ($\theta_D$) is given as

$$\theta_D = \frac{\hbar}{k_B} 3\sqrt{n * 6\pi^2 \sqrt{V}} \sqrt{\frac{B}{M} f_v} \quad (8)$$

where $k_B$ is the Boltzmann constant, $\hbar$ is Planck's constant, and n is the number of atoms in the primitive cell with volume V unit. A function of the Poisson ratio $v$, denoted as $f_v$, and M, the compound's mass, correspond to V. For CoVAs HH alloy, the predicted Debye temperature ($\theta_D$) at 10 GPa pressure is 1150.056 K. One important metric for characterizing a material's brittleness or ductility is the Pugh ratio ($\frac{B}{G}$). In general, brittle materials have a ($\frac{B}{G}$) ratio smaller than 1.75. The calculated ($\frac{B}{G}$) ratio values for CoVAs HH alloy show that the alloy is brittle in nature. In addition to this many other parameters for CoVAs HH alloy has been computed and listed in Table-1

### 3.2. Electronic and Magnetic properties

The electronic and magnetic properties of the most stable ferromagnetic (FM) $\alpha$ phase in the $C1_b$ structure is conducted through density of states (DOS) and band structure calculations. Building on the previous section, the DOS and band structures for CoVAs, a 19-valence electron alloy, are computed utilizing the modified Becke-Jones (mBj) exchange correlation potential, renowned for its accuracy in predicting band strucutres and DOS. The spin-polarised total and projected DOS for CoVAs, visualized in figure-2(a) and figure 2(b), reveal a striking asymmetric distribution of majority and minority spin states at the Fermi energy $E_F$, unequivocally confirming the alloy's magnetic nature.

**Table 3**
Elastic constants ($C_{ij}$), Bulk, Shear and Young Modulus ($B_V$, $B_R$, B, $G_V$, $G_R$, G and $E_V$, $E_R$, E in GPa), Reuss and Hill Poisson's coefficient($v_V$, $v_R$ in GPa), Kleinman's parameter ($\zeta$), Transverse, Longitudnal and Average wave velocity ($V_t$, $V_l$ and $V_a$ in m/s), Debye Temperature ($\theta_D$ in K), Pugh's Ratio (k), Chen and Tian Vickers hardness ($H^{C_V}$, $H^{T_V}$ in GPa), Lame's first and second parameter ($\lambda$, $\mu$ in GPa) of CoVAs HH alloys under external stress of 0, 5 and 10 GPa.

| Alloy | CoVAs | | |
|---|---|---|---|
| Stress | 0GPa | 5Gpa | 10GPa |
| Parameters | | | |
| $C_{11}$ | 840.079 | 907.471 | 942.093 |
| $C_{12}$ | -208.498 | -139.077 | -187.316 |
| $C_{11}$-$C_{12}$ | 1048.578 | 1046.548 | 1129.409 |
| $C_{11}$+2$C_{12}$ | 423.082 | 629.317 | 567.461 |
| $C_{44}$ | 538.722 | 573.875 | 560.603 |
| $B_V$ | 141.028 | 209.772 | 189.154 |
| $B_R$ | 141.028 | 209.772 | 189.154 |
| B | 141.028 | 209.772 | 189.154 |
| $G_V$ | 532.949 | 553.635 | 562.244 |
| $G_R$ | 532.949 | 552.504 | 562.236 |
| G | 532.949 | 553.069 | 562.240 |
| $E_V$ | 707.555 | 883.582 | 847.260 |
| $E_R$ | 707.555 | 882.621 | 29.250 |
| E | 707.555 | 883.102 | 847.255 |
| $v_V$ | -0.336 | -0.202 | -0.247 |
| $v_R$ | -0.336 | -0.202 | -0.2470 |
| $v$ | -0.336 | -0.202 | -0.247 |
| $\zeta$ | -0.131 | -0.031 | -0.080 |
| $V_t$ | 8587.320 | 8635.315 | 8610.926 |
| $V_l$ | 10855.319 | 11300.790 | 11126.976 |
| $V_a$ | 9131.428 | 9243.207 | 9195.490 |
| $\theta_D$ | 1123.878 | 1147.538 | 1150.056 |
| k | 0.265 | 0.379 | 0.336 |
| $H^{C_V}$ | 369.914 | 247.128 | 287.582 |
| $H^{T_V}$ | 355.379 | 242.320 | 280.967 |
| $\lambda$ | -214.240 | -158.941 | -185.673 |
| $\mu$ | 532.901 | 553.069 | 562.240 |

The alloy's majority spin channel exhibits metallic behavior, evidenced by the conduction and valence band overlap at the Fermi level ($E_F$). In contrast, for the minority spin channel, the total DOS exhibits a distinctive pair of bonding and antibonding states, emergent from the hybridization of higher valence Co-d and lower valence V-d states, separated by a notable 1.174 eV semiconducting band gap at the Fermi level. This confirmed the half metallic nature of the alloy.

Additionally, the As states are situated below the bonding states, demarcated by a pronounced p-d gap. The semiconducting nature of CoVAs can be attributed to the electron filling of the system, wherein the 19 valence electrons are accommodated within the available As-s, As-p and Co-d-V-d hybridized bonding states. A closer examination of the figure 2(a) confirmed that Co-d and V-d projected DOS reveals that the bonding states above Fermi level $E_F$ derive primarily from the 3d states of Co, while the antibonding states are predominantly composed of V-d states. In contrast below Fermi level $E_F$ Co-d and V-d states switched the roles. Upon closer examination of figure-2(b) reveals that the Co and V hybridization and bonding states gives rise to a fascinating splitting of Co-d-states into threefold degenerate $t2_g$ states and twofold degenerate $e_g$ states(represented with dotted lines), with the latter positioned lower in enegy as shown in figure. This nuanced understanding of the DOS structure of CoVAs provides valuable insights into origin of magnetic and semiconducting nature.

From figure 2(c), the band structure analysis for the majority spin channel reveals a metallic character (totally in aggreement with DOS analysis), characterized by the fusion of valence and conduction bands. Figure-2(d) presents the band structure for minority spin channel of CoVAs around the Fermi level, revealing crucial insights into splitting of states. Notably, an indirect band gap of 1.174 eV is observed, separating the valence band (at $\Gamma$ center) and conduction band (at $X$ center). This finding is in excellent aggreement with the DOS analysis, providing a comprahensive insights into semiconducting nature of the alloy under study. Due to the tetrahedral structural network at the $\Gamma$ center gives rise to a fascinating splitting of the top of the valence band, comprised of Co-V d-states. Specifically, this splitting results in threefold degenerate $t2_g$ states and twofold degenerate $e_g$ states, corroborating the findings of the DOS analysis.

As expected, the total magnetic moment in the ferromagnetic (FM) $C1_b$ phase of $CoVAs$, calculated to be 1.0000 $\mu_B$ (mBj), predominantly arises froom the V atoms (0.8535 $\mu_B$, while Co atom contribute moderately (0.0835 $\mu_B$), and As atom exhibit a negligible negative magnetic moment (-0.0034 $\mu_B$). Table-2 presents the total magnetic moment of the alloy and the individual magnetic moments of its constituent atoms. Notable, the Slater Pauling (SP) rule [45] is satisfied in CoVAs , where the total spin magnetic moment (M) is given by M = ($N_V$-18) $\mu_B$, with $N_V$ being the number of valence electrons per unit cell. With 19 valence electrons per unit cell in CoVAs, the calculated total magnetic moment aligns perfectly with the SP rule. The observed agreement substantiates the feromagnetic ordering, half-metallic behavior with complete spin polarization, and inherent Heusler alloy nature of the alloy under investigation.

### 3.3. Thermodynamic propeties

We have extensively investigated the thermodynamic response of novel CoVAs HH alloy is in the domain of quasi-harmonic Debye approximation model as implemented in Gibbs2 simulation package. The different parameters which reflect the thermodynamic stability of proposed alloy are examined over a temperature range of 0 to 600 K and pressures up to 10 GPa. This approach enables a thorough investigation of key thermodynamic parameters such as the volume versus temperature behaviour, Debye temperature, heat capacity, thermal expansion coefficient and entropy of the given material. By knowing these properties, we investigated the thermodynamic behavior and characteristics of these materials, offering important insights into their structural stability and performance under different temperature and pressure conditions. We calculated the volume variation shown in Figure 3(a), where a significant change occurs with an increase in temperature and a significant decrease is observed with pressure. This makes it evident that pressure plays a major role in determining the extent of occurrence and general nature of thermophysical properties.

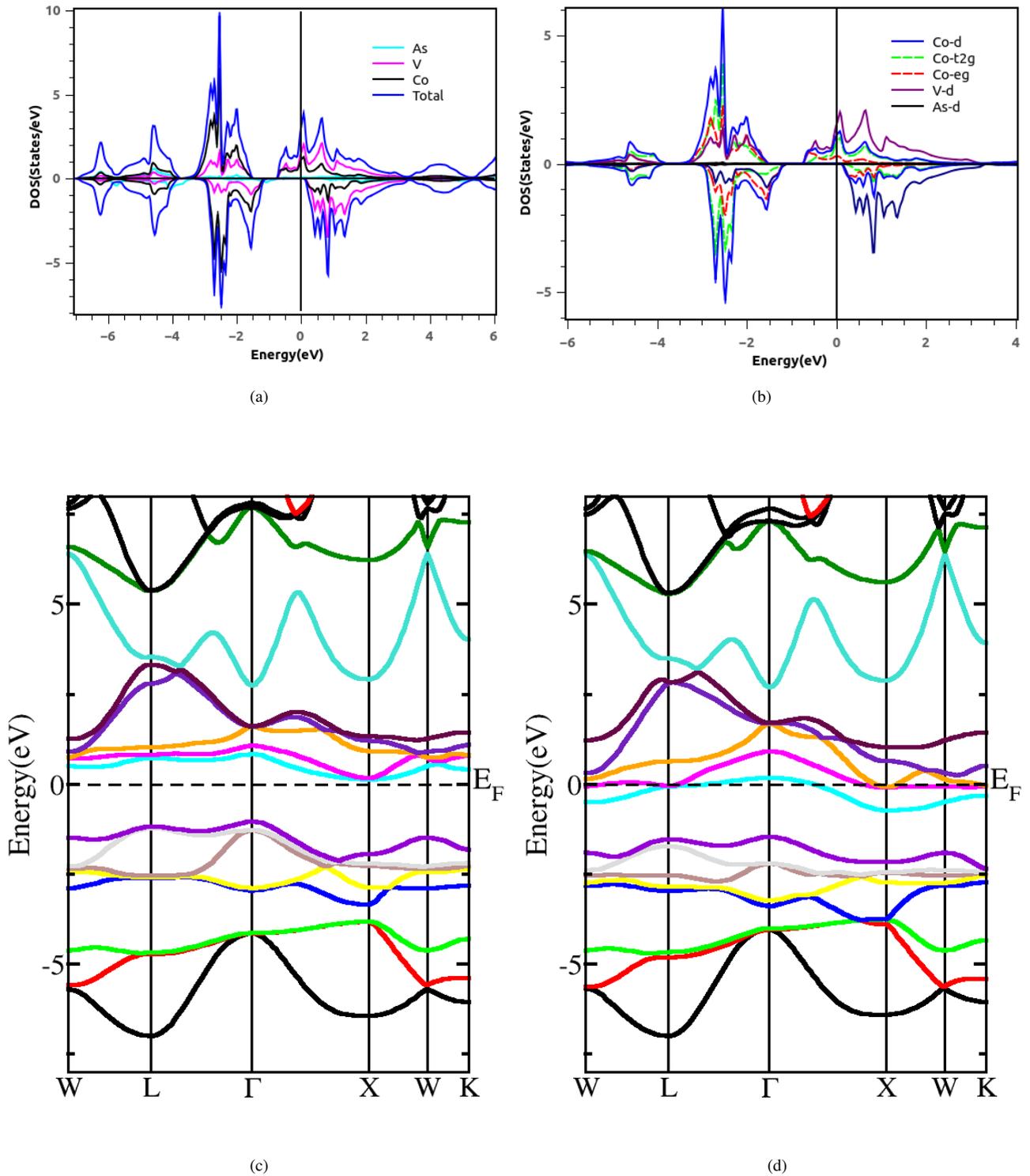

**Figure 2:** Calculated DOS and bandstruture for CoVAs using mBj (a) Total DOS (b) Partial DOS (c)spin down channel (d) spin up channel

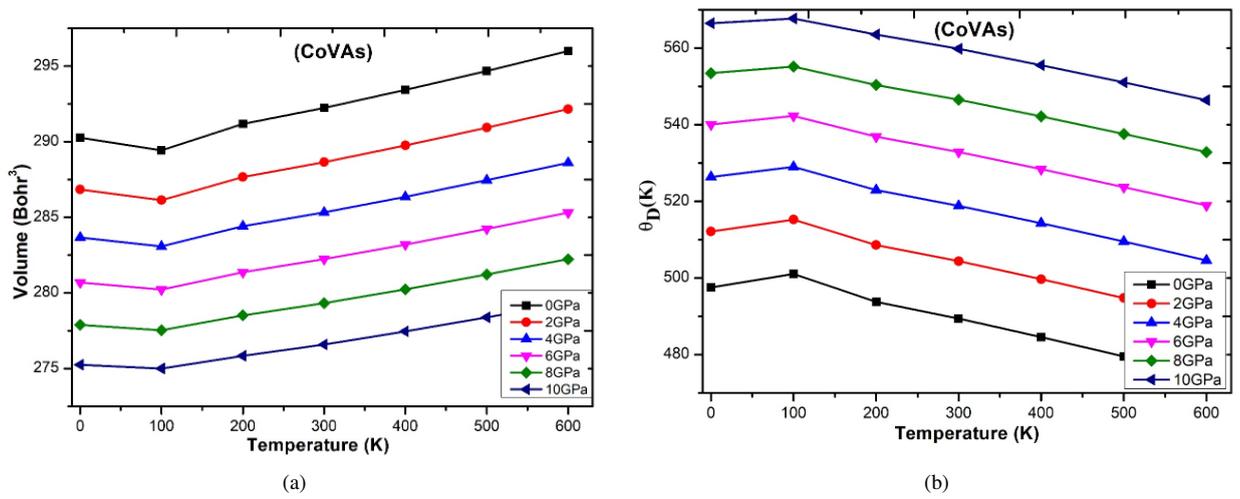

**Figure 3:** [a] Variation of Vol vs T and [b] Variation of $\theta_D$ vs T for CoVAs HH alloy.

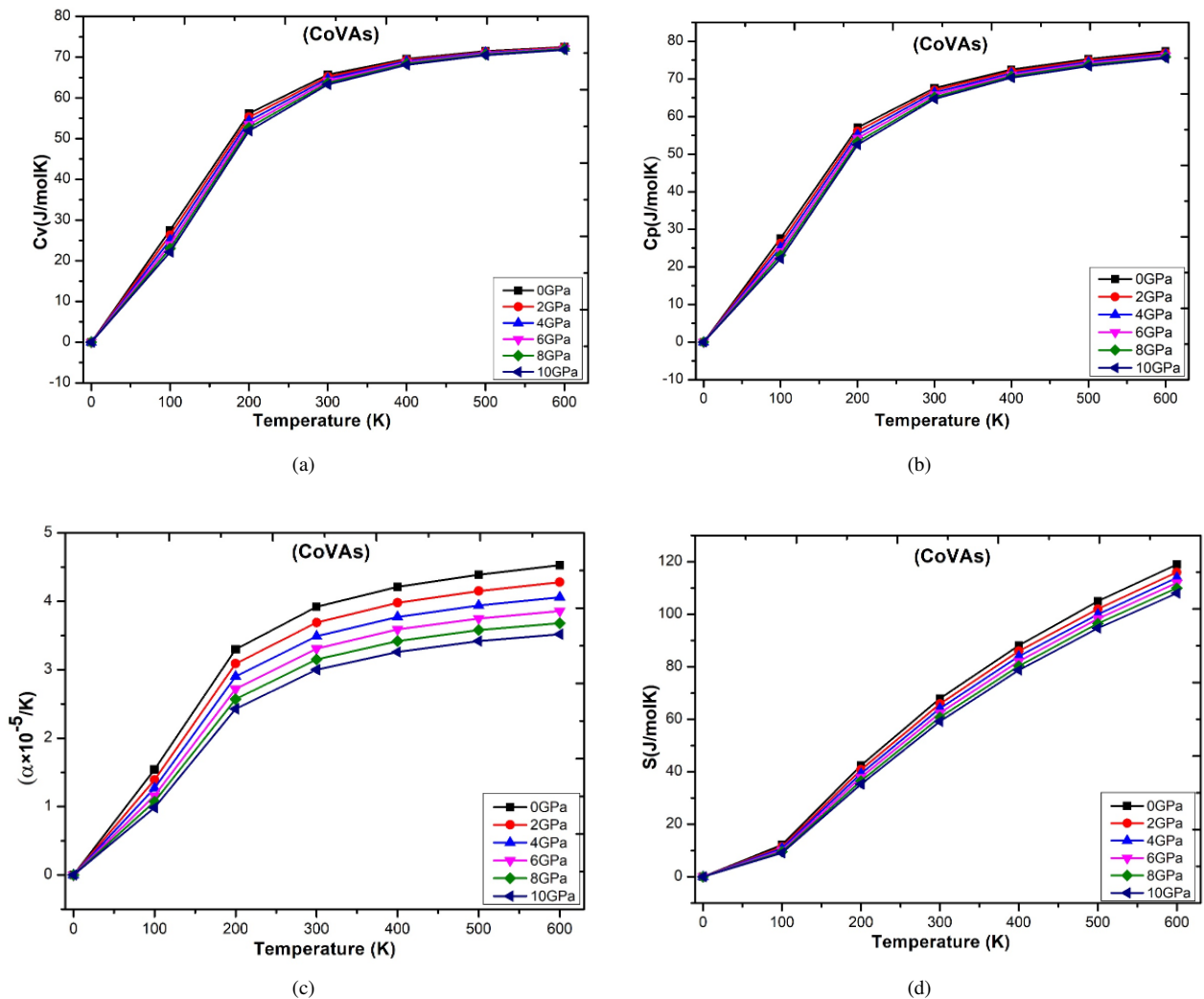

**Figure 4:** [a] Variation of Vol vs T and [b] Variation of $\theta_D$ vs T [c] Variation of $\alpha$ vs T and [d] Variation of entropy(S) vs T for CoVAs HH alloy.

**Table 4**
Total and Partial magnetic moments of $CoVAs$

| Magnetic moment per unit cell $\mu_B$ | Magnetic moment ($\mu_B$) |
|---|---|
| $\mu_{tot}$ | 1.0000 (mBJ) 0.9999 (GGA) |
| $\mu_{Co}$ | 0.0835 (mBJ) -0.1101(GGA) |
| $\mu_V$ | 0.8535 (mBJ) 1.0082 (GGA) |
| $\mu_{As}$ | -0.0034 (mBJ) -0.0252 (GGA) |
| Interstitial | 0.0664(mBJ) 0.1270 (GGA) |

The computed Debye temperature, a crucial factor influencing several physical properties of material, such as specific heat and thermal expansion, is shown in Figure 3(b). The novel CoVAs HH alloy have Debye temperatures of 490.90 K at room temperature and 0GPa pressure, respectively. Notably, the Debye temperature displays a decreasing tendency with growing temperature, while it increases with increased applied pressure. Furthermore, pressure has a greater impact on the Debye temperature than temperature, suggesting that it has a major impact on alloy's thermal behavior. This finding emphasizes how crucial it is to comprehend pressure-temperature connections in order to clarify the thermodynamic characteristics of solids. The given Figure 4(a) and 4(b) shows how heat capacity ($C_V$ and $C_P$) changes with temperature. As the temperature rises, the $C_V$ and $C_P$ increases; this is especially evident below 300K. The fact that this rise slows down at higher temperatures, however, indicates that the ion-to-ion interactions in novel CoVAs HH alloy have a major impact on heat capacity at lower temperatures. The $C_V$ and $C_P$ curves exhibit a linear relationship with the Dulong-Petit limit at high temperatures. Notably, temperature and pressure have opposing effects on heat capacity values; for pressures lower than 300 K, temperature has a greater effect. Higher $C_V$ and $C_P$ values for CoVAs system indicate their stronger heat-absorbing or heat-releasing capabilities. The extensive behaviour of material under study places it in the category of novel alloy for various thermodynamic applications and consequently has important ramifications for thermal controls.

Figure 4(c) displays the variation of the thermal expansion coefficient with temperature. Interestingly, the impact of pressure is less noticeable between 0 and 200 K. The curves show a softer slope as the temperature increases, indicating that thermal expansion is mostly influenced by temperature at lower temperatures. Furthermore, at a steady temperature, the thermal expansion coefficient falls as pressure rises. This finding emphasizes how temperature and pressure interact intricately to determine a material's thermal expansion behavior. Predicting how the alloy under investigation will react to changes in environmental circumstances, especially in situations where temperature and pressure variations are frequent, requires an understanding of these dynamics. Because of the suppression of anharmonic effects, $\alpha$ increases quickly at lower temperatures and eventually becomes constant at higher ranges. From 0 to 10 GPa, the magnitude of progressively decreases with increase in the pressure. Figure 4(d) demonstrates the entropy (S) and temperature dependence for CoVAs alloy at different pressure. Entropy values increase with temperature and decrease with pressure. From the figure it is evidemt that, entropy increases quickly at lower temperatures, whereas a slight fluctuation in entropy is evident at higher temperatures. This study makes the fact clearer that entropy is more sensitive to temperature than pressure.

### 3.4. Optical properties

The optical properties are crucial for understanding the electronic strucutre and enable the alloy's use in a wide range of technological applications, including solar energy conversion, photodetectors, optoelectronics, magneto-optical storage, and plasmonic devices. The tunable band gaps, high reflectivity, and magneto-optical effects of Cobalt based Heusler alloy make them particularly attractive for advances material research. Fig. 5(a) illustrates the absorption coefficient for the alloy under investigation. In addition to describing the quantity of light energy that semiconductors absorb, the absorption coefficient also displays abrupt absorption edges, which are crucial for determining the band gap when obstructing energy exceeds it. Electrons cannot move from the valence to the conduction band due to the incident energy below the band gap. Surprisingly, no absorption happened when there were no photons on the surfaces of any of this composition. However, when plotted against the incident photon frequency range of 0-14 eV, the absorption of CoVAs rose dramatically with increasing incident radiation frequency up to 14 eV.

The displayed graphs show that the CoVAs HH alloy has the highest absorption coefficient values, which are about $200 \times (10^4/cm)$. These materials can absorb a wide range of UV and electromagnetic radiations, according to the predicted optical absorption values, making them appropriate for optoelectronic applications. The ability of an outer layer to reflect light is known as reflectivity. Fig. 5(b) illustrates the relationship between reflectivity and photon energy. Photon energy achieves its maximum reflectivity value at 13.0 eV. For CoVAs, a static R(0) value of 0.65 has been noted. The energy loss function L($\omega$) is another optical parameter which may be defined as the energy lost by dispersion or scattering during an electron's transition.

The correlation in Fig. 5(c), which plots the optical loss function vs incident photon frequency for CoVAs composition in the range of 0 to 14 eV, is based on the scattering probabilities that emerge during inner shell transitions. The plots show that for the CoVAs HH alloy, the maximal optical loss values are 0.25 at a photon energy of 10eV. According to the energy loss spectrum, the solid reacts when an external electromagnetic disturbance is introduced. Materials that have been found to have the highest absorption in the UV region, where there is less energy loss, reflection, and dispersion, can be used to harvest optical energy for use in optoelectronics applications. The optical current caused by free carriers created as a result of incident energy is represented by the electrical conductivity, which is denoted by $\sigma[\omega]$.

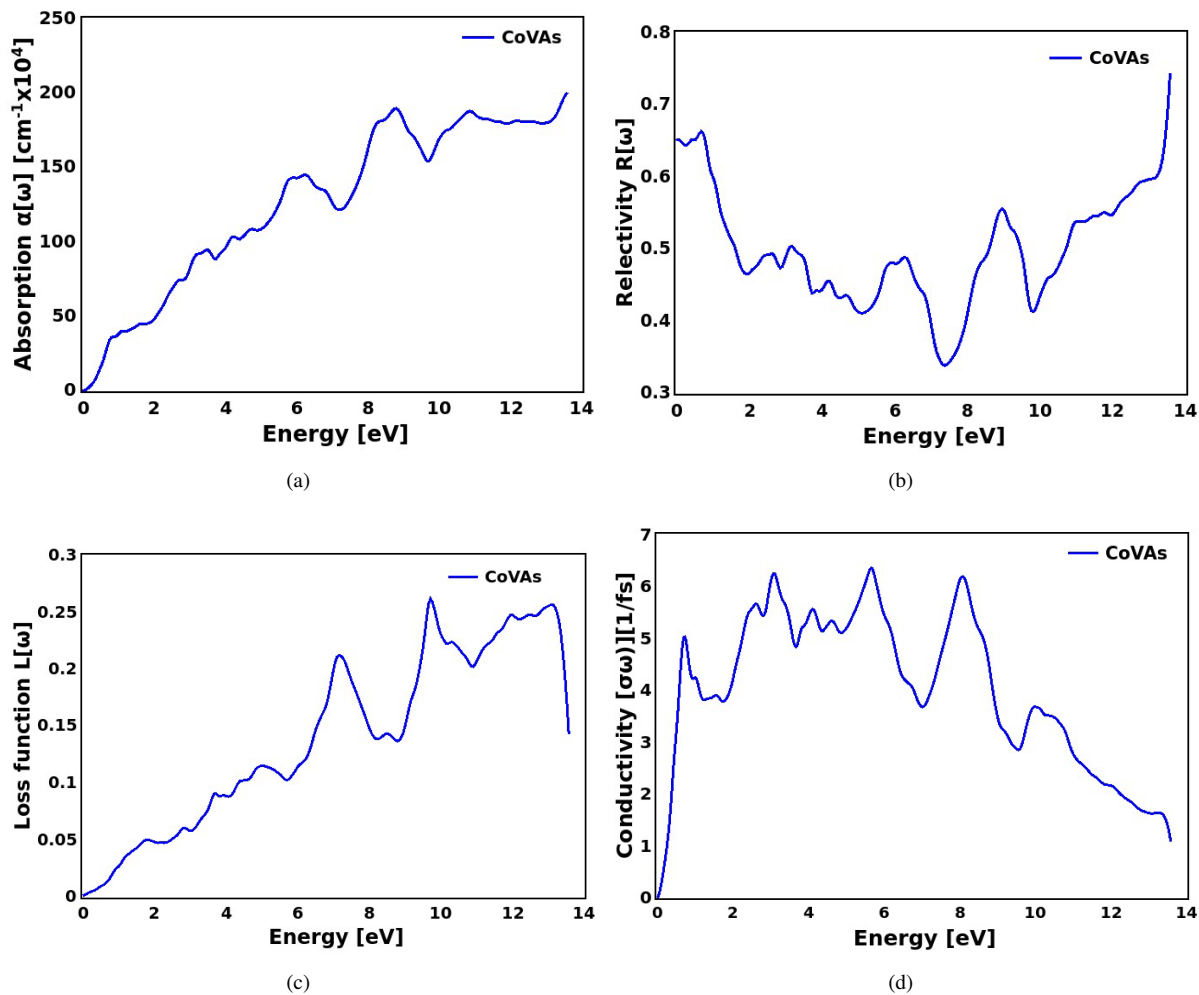

**Figure 5:** Optical properties for CoVAs HH alloy [a]Absorption coefficient, [b]Reflectivity, [c] Loss function, [d] Conductivity

Bound electrons in the valence region are excited by incoming photon energy and migrate to the conduction band. Fig. 5(d) shows the displacement of the optical conductivity versus energy graph. At 6 eV photon energy, the highest conductivity value measured for the CoVAs HH alloy is 6.2 (1/fs). The optical examination of CoVAs composition indicates that it is an excellent material for optoelectronic applications because of its excellent photon conductivity at low energy. The many peaks found in the absorption spectra indicate the electronic transitions and interactions between the different energy states. They also show the intricate relationships within the electronic structure of the substance under study.

### 3.5. Lattice dynamics studies

Phonon dispersion in the Brillion zone, illustrated in figure 6(a), were analyzed to assess the dynamical stability of the alloy under investigation. The dispersion relations were calculated using the DFPT method implemented in the phonopy package [46]. We can accurately forecast a material's thermodynamic stability by closely scrutinising its phonon band structure. As opposed to the optical modes, which are characterised by low dispersion and group velocity, the acoustic phonons, which have a high degree of dispersion, are the main contributors to thermal current in the lattice. Absence of any negative frequency in the dispersion plot clearly establish the stability of alloy under investigation. Some materials, such as thermoelectric and thermal insulators, are widely used in a variety of industrial and commercial applications. For best results, they need very low lattice thermal conductivity. We have also computed figure of merit for CoVAs HH alloys as shown in fig. 6(b). The figure of merit (ZT) is a basic property for materials that

CoVAs HH alloys for energy applications

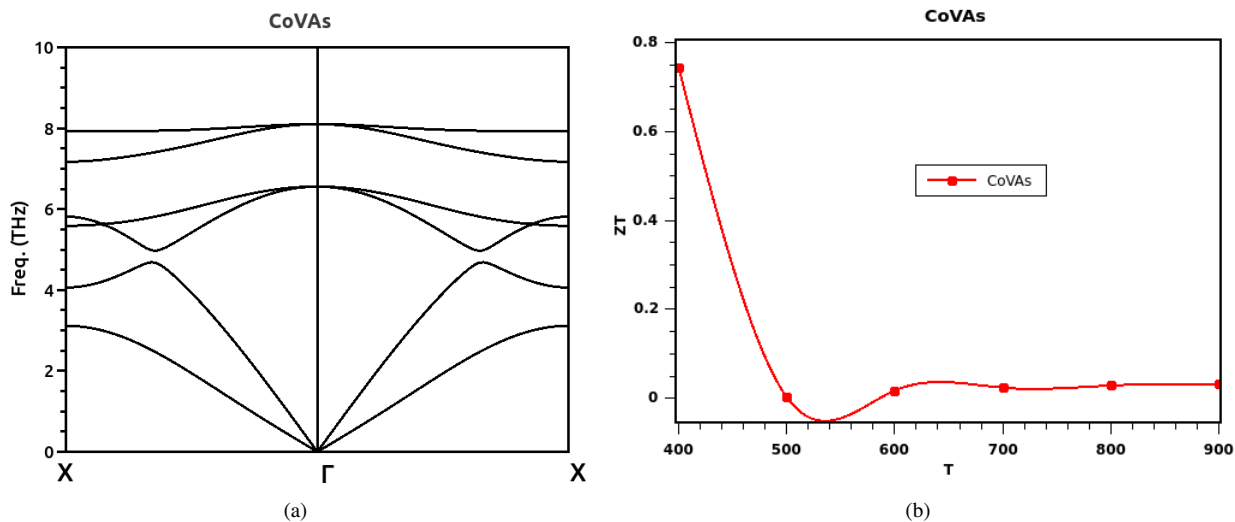

**Figure 6:** Plots of [a] phonon dynamic stability and [b] ZT for CoVAs HH alloy

is used to assess their thermoelectric effectiveness. Greater ZT values imply thermoelectric properties that are more effective. The thermal performance (ZT) of the devices is expressed as follows:

$$ZT = S^2 \frac{\sigma T}{K} \quad (9)$$

The values of ZT is 0.74 at 400 K and is decreasing with an increase in temperature as depicted in fig.

## 4. Conclusion

In this study, we have systematically investigated the structural, mechanical, electronic, magnetic, thermodynamic, and optical properties of the novel CoVAs half-Heusler (HH) alloy using first-principles calculations. Our findings reveal that among the three possible structural configurations, the $\alpha$ type phase is the most energetically stable in the ferromagnetic (FM) phase. The enthalpy of formation, and elastic constants further support the robust stability of this alloy. The electronic structure analysis confirmed that CoVAs exhibits half-metallic behavior with a substantial semiconducting band gap of 1.174 eV in the minority spin channel, while maintaining a metallic nature in the majority spin channel. The DOS analysis further corroborates this, demonstrating the hybridization of Co-d and V-d states near the Fermi level, which is responsible for its half-metallic nature. The calculated total magnetic moment of 1.0000 $\mu_B$ aligns perfectly with the Slater-Pauling rule, conirming its half-metallic ferromagnetic nature with 100 % spin-polarization at Fermi level. The thermodynamics investigations of the alloy confirmed the thermal stability and heat capacity behavior of the alloy under varying temperature and pressure conditions. The decreasing trend in Debye temperature with increasing temperature and its rise with pressure suggest strong thermal resistance, making it suitable for high-temperature applications, The heat capacity and entropy trends indicate a stable thermodynamic response, while the thermal expansion coefficient analysis confirms the minimal impact of pressure at lower temperature. Furthermore, the optical properties, specifically the absorption coefficient and refelctivity spectra, illustrates the alloys's strong interaction with incident photons, highlighting its potential for optoelectronic applications. The sharp absorption edges align with the computed band structure, reinforcing its semiconducting nature in one spin channel and metallic behavior in the other. Phonon dispersion relations without negative frequencies confirm the compound's stability and feasibility for experiment synthesis. The figure of merit (ZT = 0.74) validates the alloy's suitability for thermodynamic applications. Overall, the CoVAs HH alloy demonstrates remarkable stability, half-metallicity, significant spin polarization, and excellent thermodynamic and optical properties, making it a promising candidate for spintronics, photovoltaic, solar energy conversion, photodetector, optoelectronic and nagneto-optics storage applications. Further experimental validations and theoretical studies could provide deeper insights into its potential for real-world technological advancements

## 5. Declaration of Competing Interest

The authors declare that they have no known competing financial interests or personal relationships that could have appeared to influence the work reported in this paper.

## 6. Acknowledgement

Authors gratefully acknowledge the support by Department of Higher Education, Government of Himachal Pradesh, Shimla (Himachal-Pradesh).

# CRediT authorship contribution statement

**Sumit Kumar:** Software, Workstation, Data generation, Conceptualization, Methodology, Data curation, Writing - original draft Writing - review and editing. **Diwaker:** Conceptualization, Methodology, Data curation, review and editing. **Ashwani Kumar:** software, workstation, draft - review . **Vivek:** software, workstation. **Arvind Sharma:** draft - review. **Karan S. Vinayak:** Data curation, review and editing original draft. **Shyam Lal Gupta:** Conceptualization, Methodology, review and editing original draft.